\documentclass[journal]{IEEEtran}

\usepackage{colortbl}
\usepackage[pdftex]{graphicx}

\ifCLASSINFOpdf

\else

\fi

\usepackage[cmex10]{amsmath}

\usepackage{amsmath, amsthm, amssymb, amsfonts, array}
\usepackage{graphicx}
\usepackage{caption}
\usepackage{subcaption}
\usepackage[normalem]{ulem}
\usepackage{xfrac}

\usepackage{lipsum}% http://ctan.org/pkg/lipsum
\usepackage{multicol}% http://ctan.org/pkg/multicols
\usepackage{graphicx}

\usepackage[para,online,flushleft]{threeparttable}
\usepackage{perpage}
\MakePerPage{footnote}

\begin{document}
\title{On Feasibility of 5G-Grade Dedicated RF Charging Technology for Wireless-Powered Wearables}

\author{Olga Galinina$^{\dagger}$, %~\IEEEmembership{Student~Member,~IEEE,}
        Hina Tabassum, %~\IEEEmembership{Student~Member,~IEEE,}
        Konstantin Mikhaylov, %~\IEEEmembership{Member,~IEEE,}
        Sergey Andreev, %,~\IEEEmembership{Senior~Member,~IEEE}%
        Ekram Hossain, %~\IEEEmembership{Member,~IEEE,}
        Yevgeni Koucheryavy%,~\IEEEmembership{Senior~Member,~IEEE}% <-this % stops a space
\thanks{O. Galinina, S. Andreev, and Y. Koucheryavy are with Tampere University of Technology, Finland. }% <-this % stops a space
\thanks{H. Tabassum and E. Hossain are with University of Manitoba, MB, Canada.}% <-this % stops a space
\thanks{K. Mikhaylov is with University of Oulu, Finland.}
\thanks{$^{\dagger}$O.~Galinina is the contact author: Room TG412, Korkeakoulunkatu 1, 33720, Tampere, Finland; e-mail: olga.galinina@tut.fi}}% <-this % 
%\thanks{\textbf{April 2015}; Wireless Energy; Editor: Prof. Dusit Niyato.}}

\maketitle

\begin{abstract}
For decades, wireless energy transfer and harvesting remained of focused attention in the research community, but with limited practical applications. Recently, with the development of fifth-generation (5G) mobile technology, the concept of dedicated radio-frequency (RF) charging promises to support the growing market of wearable devices. In this work, we shed light on the potential of wireless RF power transfer by elaborating upon feasible system parameters and architecture, emphasizing the basic trade-offs behind omni-directional and directional \textcolor{black}{out-of-band} energy transmission, providing system-level performance evaluation, as well as discussing open challenges on the way to sustainable wireless-powered wearables. The key aspects highlighted in this article include system operation choices, user mobility effects, impact of network and user densities, as well as regulatory issues. Ultimately, our research targets to facilitate the integration of wireless RF charging technology into the emerging 5G ecosystem.
\end{abstract}

%\begin{IEEEkeywords}
%\end{IEEEkeywords}
%-------------------------------------------------------------------------------------------------------------------------------------------------------------------%
\section{Introduction and Motivation}

\subsection{Wearables in 5G Background}

Mobile communications technology is rapidly moving towards its fifth generation (5G), which is expected to become commercially available beyond 2020. %However, as a consequence of respective research and development efforts, it is currently understood that the challenging 5G network capacity objectives can only be met with further \textit{densification} of existing cellular deployments, where a large number of small base stations (SBSs) would provide the much needed user data rates~\cite{Bhu14}. %These SBSs are typically equipped with licensed-band (e.g., 3GPP LTE) technology as well as with supplementary unlicensed-band (e.g., WiFi) radios~\cite{Gal15}, and we expect this integration to continue further by potentially including mmWave-based access solutions.
%The emerging 5G ecosystem will also embrace a rich variety of unattended wireless devices (such as actuators, sensors, and smart meters) comprising the Internet of Things (IoT). 
Within the emerging 5G ecosystem bridging across human and machine realms,
%At the intersection of human and machine realms, 
there stands a fascinating innovation, which promises to develop into a \$70-billion market by 2025~\cite{Har15}. Commonly referred to as wearable electronics or, simply, \textit{wearables}, this niche features a plethora of smart, connected companion devices, as well as a wide range of apparel and textiles. Being a decisive departure from the smartphone form-factor and functionality, contemporary wearables adopt miniaturization in communications, \textcolor{black}{sensing}, and battery technology.% and hence significantly augment human capabilities.
\begin{figure}[!ht]
\centering
\includegraphics[width=1.0\columnwidth]{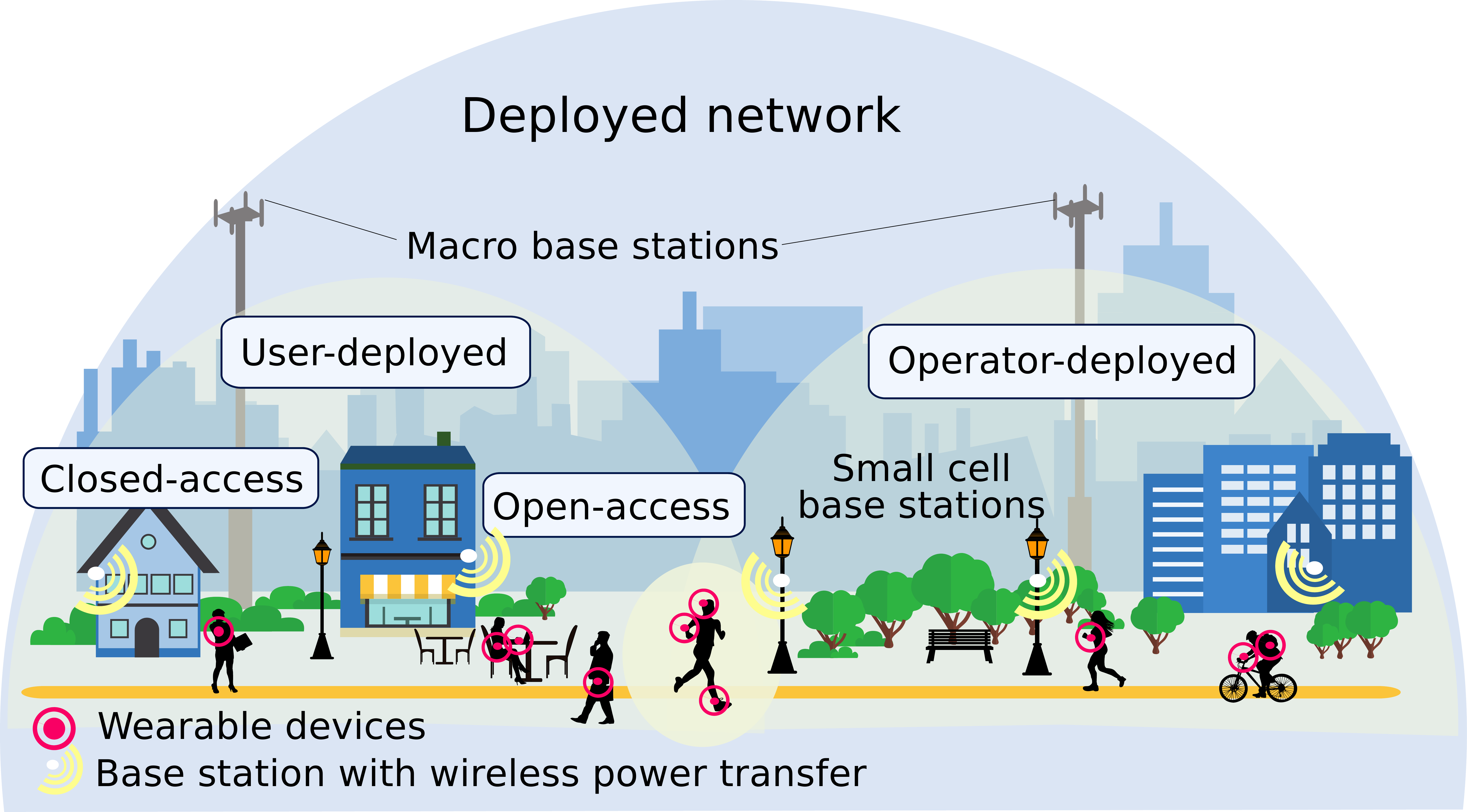}
\caption{Overall vision of 5G-ready wireless charging.}
\label{fig:main} \vspace{-20px}
\end{figure}
However, what appears as the attractive benefit of wearables, also becomes their major limitation, which may ultimately hamper the overall product adoption. Apparently, with modest expectations for near-term battery capacity improvement, the power-hungry electronics threatens to render wearable equipment an "overnight" sensation, in need of daily charging. Consequently, the fairly limited battery lifetime of current-generation wearables becomes an immediate challenge. Fortunately, there is a solution that has the potential to mitigate the imbalance between the offered and the desired operation times, which is to equip future wearables with \textit{energy harvesting} capabilities.

Broadly, energy harvesting techniques enable energy-constrained devices to replenish their charge levels without physical connections. However, harnessing the ambient solar, vibrational, thermal, chemical, biological, or electromagnetic power is inherently opportunistic, as it is only made available by intermittent energy sources. In contrast, dedicated \textit{wireless charging} across an air gap is more predictable -- it can thus help achieve the required availability and reliability of energy supply, which is becoming crucial for today's QoS-sensitive wearables (e.g., health and wellness devices, wearable cameras, and augmented reality glasses). With the considerable recent progress in energy transfer technology~\cite{Mis15}, commercial products\footnote{See, for example, commercial devices from Powercast already available for purchase, http://www.powercastco.com/} employing various forms of wireless energy harvesting and transfer are beginning to sprout on the market. \textcolor{black}{Complementing \textit{wireless-powered} wearable devices, non-wearable market also deserves attention, including that of low-energy Internet of Things} (IoT)~\cite{RFwhite}.%https://www.google.fi/url?sa=t&rct=j&q=&esrc=s&source=web&cd=1&cad=rja&uact=8&ved=0CCMQFjAAahUKEwjJ-vWPsILJAhWCjywKHRBBBI0&url=http%3A%2F%2Fwww.getfreevolt.com%2Fdownloads%2FRF%2520Energy%2520Harvesting%2520Whitepaper.pdf&usg=AFQjCNFFMySVNvxVZwfMazoxlCL7WxCJpQ&sig2=nTKn48lsmK4BBTEqIe_fvw&bvm=bv.106923889,d.bGg}.

%\footnote{White paper "Wearables: 10 Trends to Watch", Tractica, 3Q 2015}

In this article, we advocate a novel vision (see Fig.~\ref{fig:main}), where wireless-powered wearables can proactively replenish their energy supply from a large number of small base stations (SBSs) additionally equipped with a dedicated wireless energy transfer interface (WETI) operating on a separate frequency channel than the actual data transmission. % (i.e., out-of-band power transfer)}. 
 Along these lines, we detail the respective wireless charging landscape and our proposed concept in what follows. 

%\textbf{CONTRIBUTION:} In what follows, we begin with outlining the contemporary landscape behind the dedicated wireless charging technology and then conduct its in-depth feasibility analysis. Expanding on the latter, we further detail our envisioned 5G-ready system architecture and discuss the attractive deployment models of WETI-enabled small cells. Finally, we conclude with elaborating upon the new research challenges that arise from the implementation of this promising concept. The below numerical results are based on our extensive background research in the field of wireless energy harvesting and transfer, rely on the state-of-the-art technological capabilities, and employ the advanced system-level evaluation framework specifically developed for the targeted system design.

% [Lu14_2], [!Lu15], [!Lu15_2], [!Leo11], [!Tal15], [Vis13], [Ulu15], [Mis15], [Bi15], [Hua15], [Kim14]

\begin{table*}[tbp]
\caption{Comparison of various frequency bands for RF charging}
\vspace{-0.2cm}
	\centering
	%\caption{Estimation of RF energy transfer characteristics for selected bands}
	\begin{tabular}{|m{83pt}|m{14pt}|m{11pt}|m{13pt}|m{13pt}|m{13pt}|m{13pt}|m{13pt}|m{13pt}|m{13pt}|m{11pt}|m{12pt}|m{12pt}|m{14pt}|m{14pt}|m{14pt}|m{12pt}|}
		%\hline
		%Parameter & Frequency 1&Frequency 2&Frequency 3&Frequency 4&Frequency 5&....\\
		\hline
    		  Parameter\footnote{1} &\multicolumn{6}{|c|}{ISM (industrial, scientific, and medical) bands} &\multicolumn{10}{|c|}{Cellular bands} \\ % <---- inserted &
    		\hline 

		Band& \multicolumn{2}{|m{42pt}|}{915 MHz}& \multicolumn{2}{|m{42pt}|}{2.4 GHz}& \multicolumn{2}{|m{42pt}|}{5.8 GHz}& \multicolumn{2}{|m{41pt}|}{850 MHz}& \multicolumn{2}{|m{40pt}|}{1.7 GHz}& \multicolumn{2}{|m{41pt}|}{2.1 GHz}& \multicolumn{2}{|m{42pt}|}{1.9 GHz}& \multicolumn{2}{|m{42pt}|}{2.5 GHz}\\
		%5725-5875 MHz&2400-2483.5 MHz&869.4-869.65 MHz &433.05-434.79 MHz&138.2-138.45 MHz&40.66-40.79 MHz\\%26.85-27.255 MHz\\
		\hline
		Wavelength, m& \multicolumn{2}{|c|}{0.33}& \multicolumn{2}{|c|}{0.12}& \multicolumn{2}{|c|}{0.05}& \multicolumn{2}{|c|}{0.35}& \multicolumn{2}{|c|}{0.18}& \multicolumn{2}{|c|}{0.14}& \multicolumn{2}{|c|}{0.16}& \multicolumn{2}{|c|}{0.12}\\
			
		Array size\footnote{2}, m&\multicolumn{2}{|c|}{0.33x0.33}&\multicolumn{2}{|c|}{ 0.12x0.12}&\multicolumn{2}{|c|}{ 0.05x0.05}&\multicolumn{2}{|c|}{ 0.35x0.35}&\multicolumn{2}{|c|}{ 0.18x0.18}&\multicolumn{2}{|c|}{0.14x0.14 }&\multicolumn{2}{|c|}{0.16x0.16 }&\multicolumn{2}{|c|}{0.12x0.12 }\\
		\hline
		&Omni&Direct.&Omni&Direct.&Omni&Direct.&Omni&Direct.&Omni&Direct.&Omni&Direct.&Omni&Direct.&Omni&Direct.\\
		\hline
		Energy radius, m&10.57&34.85&3.95 &13.01 &1.67&5.50&11.38&37.51&5.69&18,76&4.60&15.18&5.09&16.78 &3.87 &12.75\\
		$P_{\text{harvested}}$\footnote{3}, $\mu$W&11.17&	121.44&	1.56&	16.94&	0.28&	3.02	&	12.94&	140.73&	3.24	&35.18&	2.12	&23.06&	2.59	&28.16&	1.50&	16.27\\
		%Replenishment rate\footnote{4}, \%&	223&	2429&	31&	339&	6&	60&	259&	2815&	65&	704&	42&	461&	52&	563&	30&	325\\
		Replenishment rate\footnote{4}, \%&123&2329&	-69&242&-94&	-40&		159&	2715&	-35&604&	-58&	361&	-48&	463&	-70&225\\
Energy positive range\footnote{5}, m&	14.95&	49.28& 5.58 &	18.41 &2.36&	7.77	&	16.09&	53.05&	8.04&	26.53&	6.51&	21.47&	7.20&	23.73&		5.47&		18.04\\

Support time\footnote{6}, min&	8.11&	0.43	&N/A\footnote{7}&	4.19	&N/A&N/A	&	6.30&	0.37	&N/A	&1.66&	N/A	&2.77	&N/A	&2.16&	N/A&	4.44\\
		%WET radius (omni), m & ...&...&...&....&....&....&....&....\\
		%WET radius (direct.), m & ...&...&...&....&....&....&....&....\\
		%Buffered energy, mW& ...&...&...&....&....&....\\
		%....& ...&...&...&....&....&....&....&....\\
		\hline
		
		\hline
	\end{tabular}
\begin{tablenotes}
\raggedright
\item[1]Parameters used for estimation are: transmitter antenna gain (omni-directional/directional) 2.15/14.51dBi, receiver gain 0dBi, WET efficiency 50\%, sensitivity = $-20$dB, consumed power (discharge rate) 5$\mu$W, radiated power 1/0.63W (omni-directional/directional)\\
\item[2]Assuming 3x3 square transmitter array with $\!1\!/2$ wavelength separation   \quad \quad \quad \quad   \quad \quad 
\item[3]Harvested power at reference distance 10m\\
\item[4]Energy replenishment rate at 10m =  (harvested power/consumed power $-1$)$\cdot 100\%$\quad\quad
\item[5]Maximum distance where harvested power$\geq$consumed power\\
\item[6]Time of harvesting to support $N=10$ min of autonomous work at 10m  \quad  \quad  \quad   \quad  \quad
\item[7]Not available due to feasibility constraints
\end{tablenotes}
\vspace{-0.4cm}
\end{table*}

\subsection{Wireless Charging Landscape}

As argued above, dedicated wireless charging has the potential to power the next generation of wearables, which are essentially compact, body-worn computation and sensing platforms aimed at tracking, storing, processing, and reporting important  physiological parameters, activity, and events~\cite{Tal15}. Conveniently, wireless power transfer can, in principle, utilize the same antenna that wearables already use for communication, without the need for additional transducers required for ambient non-electromagnetic energy scavenging. Therefore, the power requirements of wearables can be met without invoking extra size, weight, complexity, or cost.

Essentially, modern wireless energy transfer and harvesting technology exists in two forms, namely, near-field and far-field. The former includes \textit{non-radiative} techniques based on magnetic induction or magnetic resonance coupling between two tightly aligned coils. While non-radiative systems enjoy high energy transfer efficiency and thus become an active area for standardization (see, e.g., Qi and A4WP specifications), they only remain usable at very short distances~\cite{Lu15}. In contrast, with \textit{radiative} far-field energy transfer methods, the incoming radio-frequency (RF) signals can be converted into electric supply over longer distances (on the order of tens of meters) and in a wide range of frequencies.% (generally, from $3 \cdot 10^3$ to $3 \cdot 10^{11}$ Hertz). 

To this end, \textcolor{black}{Table I elaborates on the feasible frequency range for RF charging and suggests the attractiveness (including size, charging range, etc.) of $915$ MHz or $850$ MHz bands. These numbers have been produced for certain center frequencies and radiated powers allowed according to the U.S. FCC restrictions (e.g., 15.247.b.4), as well as based on the tentative values for other parameters given in the footnotes.} We also note that other potential bands are available, but have not been included here due to either their low efficiency or large antenna size requirements. 

Importantly, we comparatively analyze the feasibility of \textit{omni-directional} vs. \textit{directional} RF energy transfer, which employs energy/power beamforming techniques. As the latter approach makes it possible to steer directivity electronically in a controllable manner, it can further improve energy transfer efficiency without the need for extra bandwidth or increased transmit power. Accordingly, Table I confirms better results for directional RF power transmission. \textcolor{black}{Note that hereinafter we refer to the FCC restrictions on power radiation, but the corresponding numbers may be adapted for other regulatory authorities.} %Hence, we perform our below illustrative examples for both directional and omnidirectional RF transmission based on the former, which resides in the ISM bands. 

\subsection{Our Vision of 5G-Grade Wireless Charging}
%(i) coexistence of operator-deployed and client deployed SBSs lead to different incentivization/energy sharing mechanisms
%(ii) out-of-bad SWIPT allows more flexible receiver antenna architecture

Inspired by said progress in wireless power transfer technology, we envision that future 5G-grade SBSs be equipped with dedicated RF charging capability and provide next-generation wireless-powered wearables with reliable amounts of energy. 

1) The proposed \textit{integrated SBS architecture may be preferred} over installing standalone power beacons, which are in essence low-cost wireless charging stations~\cite{Tab15}. \textcolor{black}{Generally, the power beacons do not require backhauling, but %incur additional capital and operational expenditures (CAPEX/OPEX) 
have to be deployed and separately managed, whereas the SBSs take advantage of existing operator infrastructure, which is only expected to become denser on the way to 5G. }

2) Together with the operator-deployed SBSs, additional small cells may be installed by private clients. This coexistence with closed-access, user-deployed stations, however, leads to different energy sharing incentivization mechanisms. Eventually, as a \textit{mixture of operator- and user-deployed SBSs with WETI} becomes available "on every lamp-post" in urban areas, the respective densities may just appear sufficient to enable stable RF charging.

3) The integrated multi-radio SBS deployments have another inherent advantage, as electromagnetic radiation can simultaneously carry both energy and meaningful data -- the same waveform can thus enable simultaneous wireless information and power transfer (SWIPT). Hence, the concept of \textit{SWIPT brings a new dimension to the emerging 5G ecosystem}, whether in the form of in-band (energy and information share the same band) or out-of-band (orthogonal frequency bands are utilized) system. As the former typically requires more advanced receiver architectures~\cite{Lu14-2}, the out-of-band option may be preferred at the initial stages of implementation. \textcolor{black}{We note that in this work we focus solely on the out-of-band SWIPT, not considering the data transmission explicitly.}

4) As we learn from Table I, \textit{directional RF charging is more beneficial}, but it also requires tight time synchronization and accurate channel state information (CSI) for the sharper beams to enjoy increased power density~\cite{Liu15}. \textcolor{black}{Fortunately, this valuable information may become available with the 5G-grade cellular network assistance and positioning technologies, which can also supply wearable devices with the knowledge on the availability of active energy sources, their location, energy status, service cost, etc. However, at high mobility, accurate CSI acquisition for directional energy transfer may become a challenging issue in 5G wireless communications.}

5) In addition, cellular-controlled WETI operation has the potential to enable \textit{highly-predictable on-request power supply} that can minimize the associated energy costs %. %Enjoying minimal installation overheads, the dense SBS deployments with the new WETI radio 
and may thus soon become an integral part of the maturing 5G vision.

% - simultaneous charging of multiple devices is preferred (Qualcomm report)
% - directivity is possible only for MW and laser

%\textcolor{magenta}{ this norm must meet stringent radiation standards before release \par
%Going wireless adds roughly 25 percent to the charging station, a cost increase that also increases the cost of the receiver by about the same amount. }

%%------------------------------------------------------------------------------------------------------------------------------------------------
%%------------------------------------------------------------------------------------------------------------------------------------------------
%%------------------------------------------------------------------------------------------------------------------------------------------------
%%------------------------------------------------------------------------------------------------------------------------------------------------
%%------------------------------------------------------------------------------------------------------------------------------------------------

\section{RF Charging System: Architecture Choices}

%optional: The design of a WET system introduces multiple novel challenges and tradeoffs which differ greatly from the ones arising during the development of conventional data transfer systems. 
%\begin{itemize}
%\item energy = broadcast (may be many users per beam);
%\item will there be energy, which is useful for one and not for other (semantics)?
%\item energy connection may be not seamless, can be stored; short distances; 
%\item can be physically moved;
%\item interference is good?
%\end{itemize} 

\begin{figure*}[!ht]
\centering %
%\vspace{-5px}
\includegraphics[width=\textwidth]{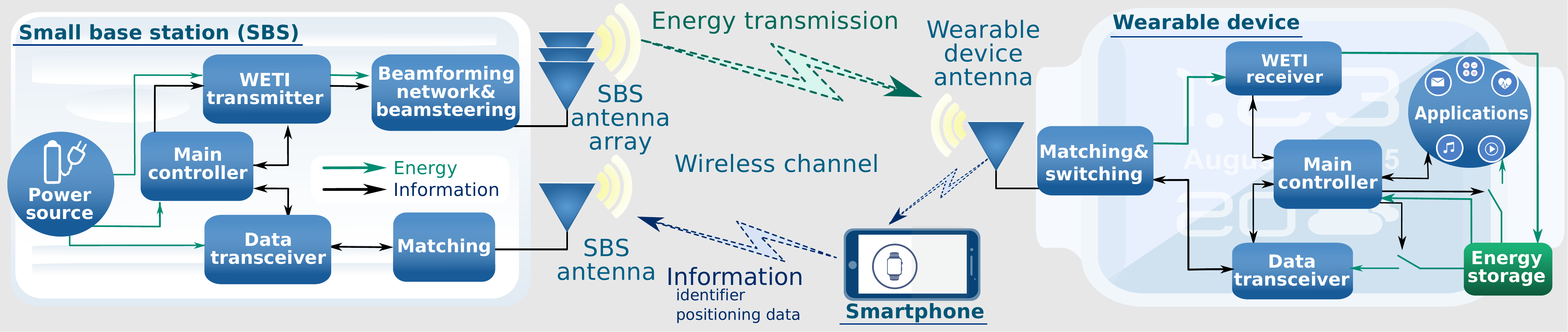}
\caption{Architecture of the envisioned RF charging system.}
\label{fig:architecture}
%\vspace{-15px}
\vspace{-0.6cm}
\end{figure*}

Our proposed wireless charging system (see Fig.~\ref{fig:architecture}) features two main components: (i) a dedicated RF power source and (ii) a wireless-powered wearable device, which acquires and converts into electricity the input wireless energy from the source. Below we discuss both components in more detail as well as clarify their intended interaction.

\vspace{-0.1cm}
%%------------------------------------------------------------------------------------------------------------------------------------------------
\subsection{RF Power Source} \label{Energy_transmission_capability}

\subsubsection{Frequency selection and waveform design} The development of an energy transmission system begins with selecting the type of the radio signal carrying energy, that is, its frequency band and waveform (single-tone, multi-tone, chaotic, etc.). In practice, a suitable frequency is chosen based on the operational environment and its expected attenuation, the local (national) RF regulations, and the restrictions on the antenna size. More specifically, dedicated RF energy transmitters may concurrently employ license-free ISM frequency bands and/or licensed operator bands that e.g., become available after the \textit{decommissioning} of the legacy cellular systems (see Table I for details). Further, the choice of the appropriate charging signal waveform is dictated by the \textit{maximum transmit power} and the \textit{power spectral density} regulations, and aims at maximizing the performance of RF-DC conversion on the client side, as well as at minimizing the complexity and energy consumption of the corresponding transmitter hardware. %In this respect, the noise-like broadband signals with high \textit{peak-to-average power ratio} are the viable choice as of today. %whereas the power source may be equipped with more complex circuitry due to more abundant resources.

\subsubsection{Omni-directional vs. directional RF charging} Depending on the target scenario, the provider of the charging service can employ two distinct alternatives: omni-directional energy transmission able to charge a large number of clients across multiple directions, or directional RF transfer serving one to several clients with its narrower power beam. While the former generally enjoys simpler implementation, exploitation, and management, the latter requires more complex procedures for antenna steering and needs additional information on the current location of a client, all of which creates signaling overhead. However, due to the lower antenna gain for omni-directional transmission, this type of charging has much shorter effective operational range and may lead to excessive power consumption. At the same time, directional power transfer provides stronger signal at more distant locations and can thus remain usable when the RF attenuation is high.

\subsubsection{Maximum radiated power restrictions} Another important aspect that affects the antenna design, as well as that of the entire SBS, is the limitation imposed by the national regulatory authorities on the maximum effective radiated power (ERP). In particular, if the antenna gain (inversely proportional to the respective beam width) is less than $6$ dBi, the ERP conducted to the antenna elements forming a single beam is constrained by $1$ W; otherwise, the maximum ERP is decreased either (i) by $1$ dB for each $3$ dBi above $6$ dBi or (ii) by $1$ dB for $1$ dBi (see FCC 15.247.b.4 for details). However, current regulatory documents have not been written with wireless power transfer in mind, and hence we report data from the existing specifications for information transmission. As a result, the ultimate effectiveness of the envisaged charging system significantly depends on the future path taken by official documentation, and there is an urgent demand to develop respective legal basis.

\subsection{Wireless-Powered Wearable}

\subsubsection{RF signal strength considerations} Unlike the integrated SBSs, modern wearables are radically constrained by their linear dimensions, weight, and cost. Therefore, a major issue in the design of an energy harvesting-capable wearable device is to assure the reception of a sufficiently strong RF signal to make wireless charging practical. To this end, a rectifying antenna (rectenna) is employed, which utilizes a diode-based circuit to convert RF signals to DC voltage. The severe attenuation of a radio signal, the need for maintaining the received power above the sensitivity threshold (to enable RF-DC conversion), as well as the unpredictable mobility of wireless clients (causing sudden changes in the orientation of the receiving antenna) -- these are just a few factors that need to be addressed. 

\subsubsection{RF-DC conversion efficiency} Another critical aspect of wireless charging is to ensure the sufficient efficiency of converting the received RF signal into the useful energy. Along these lines, the following factors have to be taken into account: (i) the need for impedance matching between the antenna and the input of the rectifier, as well as between the rectifier's output and the energy storage unit; (ii) the high variation in the RF-DC conversion efficiency in conventional diode-based rectifiers with a fixed number of stages, which is a function of frequency, waveform, and input RF power. Even though various techniques and mechanisms have been proposed to address these glaring demands, many of them remain overly complex and costly to be used in contemporary wearable devices~\cite{Vis13}. 

%The most crucial issue, however, is the actually available amount of received energy at the wireless-powered wearable, which depends on a number of parameters, such as propagation conditions, distance from the energy source, etc. %Given that wireless charging system performance remains largely location-dependent, mobile wearable devices with RF energy harvesting capability have to proactively replenish their batteries. 
\subsubsection{Rectifier sensitivity constraints} Broadly, the amount of acquired energy depends on the emitted power, wavelength, antenna gain, rectifier and its RF-DC conversion efficiency, as well as on the location-dependent factors (including path loss, energy dissipation, shadowing, scattering, and fading). Under the corresponding limitations on the radiated power, in many cases the said amount may be characterized by the Friis transmission equation:
\begin{equation}
\frac{p_{rx}(d)}{p_{tx}} = \frac{G_{tx} G_{rx} \lambda^2}{(4\pi d)^2}, \label{eqn:friis}
\end{equation}
where $\lambda$ is the wavelength, $p_{tx}$ and $p_{rx}(d)$ are the transmit and receive powers at the separation distance $d$, respectively, $G_{tx}$ and $G_{rx}$ are the transmitting and receiving antenna gains. Therefore, the sensitivity of the rectifier (determined by the minimum amount of the received power to enable energy acquisition) defines the energy transfer radius with respect to the ERP restrictions and the associated antenna gains. 

\textcolor{black}{We note that the above equation is known to hold for free-space propagation and in our preliminary calculations we further assume no obstacles or reflections at the shorter distances, where RF harvesting is possible. However, for non line-of-sight scenarios (including indoor or dense urban use cases), either the formula (\ref{eqn:friis}) could be adapted (e.g., according to the 3GPP specifications) or the corresponding statistical model may be introduced (as e.g., in \cite{obayashi1998} for body-shadowing effects).}

\subsubsection{Energy acquisition aspects} Whenever the received power exceeds the rectifier sensitivity threshold, wireless energy is acquired and accumulated for future use in the energy storage unit of a suitable capacity (which constitutes a research question on its own). Throughout their operation, wearables follow a particular power consumption profile, which may change over time as driven by the duty cycle and storing efficiency constraints (including matching and leakage). If the energy level drops to near zero, we say that the device is experiencing an \textit{outage} at the moment. In order to avoid outages, wearables may inform the charging SBS about their battery level by taking advantage of suitable network assistance protocols, so that recharging might begin after hitting a given battery charge level. 

It can thus be concluded that, given the coverage radius of the energy source and the architecture specifics of wearables, the required RF power for effective wireless charging can be determined. \textcolor{black}{We also note here that wearables may use the smartphone as a feedback channel, as well as a localization and authentication intermediary.}
%Based on the above, within the resulting coverage radius one may evaluate the amount of collected power, analyze the fraction of time spent in the charging zone, and thereby define the range of the radiated RF power that would enable effective energy collection.

%%------------------------------------------------------------------------------------------------------------------------------------------------
%%------------------------------------------------------------------------------------------------------------------------------------------------
%%------------------------------------------------------------------------------------------------------------------------------------------------

\subsection {System-Level Aspects}

%In addition, the density of power sources and users as well as the distance between them has to be accounted for. 

Above, we have addressed the individual features of the energy source and its wearable consumer. However, their respective operation and interactions become increasingly complex as we move on to considering multiple transmitters and/or receivers.

\subsubsection{Operation of several transmitters} If multiple wireless energy sources are active, it may happen that a certain wearable device can concurrently receive RF energy from several different SBSs at a time. This emphasizes the importance of selecting appropriate waveforms, which would not cause reduction in energy transfer efficiency in case of their superposition and regardless of the actual signal phase difference. For the directional transmission, the overlap of energy beams at the receiver \textcolor{black}{(causing wave interference)} may be avoided by employing intelligent signal processing and network-assisted user tracking mechanisms. %However, even 5G-grade positioning services may lack accuracy, especially in dense urban environments, which brings along the need for over-provisioning by, potentially, forming wider beams.

%%------------------------------------------------------------------------------------------------------------------------------------------------
\subsubsection{Operation of several receivers} In the RF charging system with multiple receivers, the energy source may form several dedicated beams to serve its clients (i.e., users with multiple wearables). However, from the regulatory documentation %(see e.g., FCC 15.247.b.4) 
it follows that the aggregate power transmitted simultaneously on all beams should not exceed the limit for a single beam by more than $8$ dB, which allows \textcolor{black}{for a maximum of $\lfloor 10^{8/10} \rfloor= 6$} individual beams to operate simultaneously at the highest permitted power. Moreover, if transmitted beams overlap, the power shall be reduced even further to ensure that their aggregate ERP does not exceed the upper limit set for a single beam.

\subsubsection{Number of beams and blockage} %Given the tight regulatory limitations on radiated power for the directional RF transmission, for multiple receivers it may be more efficient to use \textcolor{black}{few narrow beams, which would periodically change their direction, rather than create a multitude of simultaneous low-energy wider beams}. 
\textcolor{black}{Given the tight regulatory limitations on radiated power for the directional RF transmission, for multiple receivers it may be efficient to use beams, which would periodically change their direction.} 
Another aspect of directional energy transfer is the possibility to form so-called "null" beams, thus restricting the radiated power in certain directions. Finally, the system with many receivers has to additionally account for shadowing and other detrimental effects caused by the body (see, e.g., \cite{obayashi1998}), such as \textit{self-blockage}, when the receiver is radio-"shadowed" by the user body (parts) during motion, and \textit{blockage}, when another body accidentally intersects the energy path for the target user.% who hence occurs in a radio "shadow".

In the remainder of this text, we conduct a detailed investigation of omni-directional vs. directional RF charging technology in a characteristic 5G scenario, with a particular emphasis on user mobility. Even though the effects of mobility can in principle be controlled with network-assisted wireless charging, the number of respective studies is very limited~\cite{Lu15-2} and we bridge the indicated gap.

%------------------------------------------------------------------------------------------------------------------------------------------------------------------------------------------------------------------------------------------------------------------------------------------------------------------------------------------------------------------------------

\begin{figure*}%[!ht]
\centering
\vspace{-20px}
\includegraphics[width=0.9\textwidth]{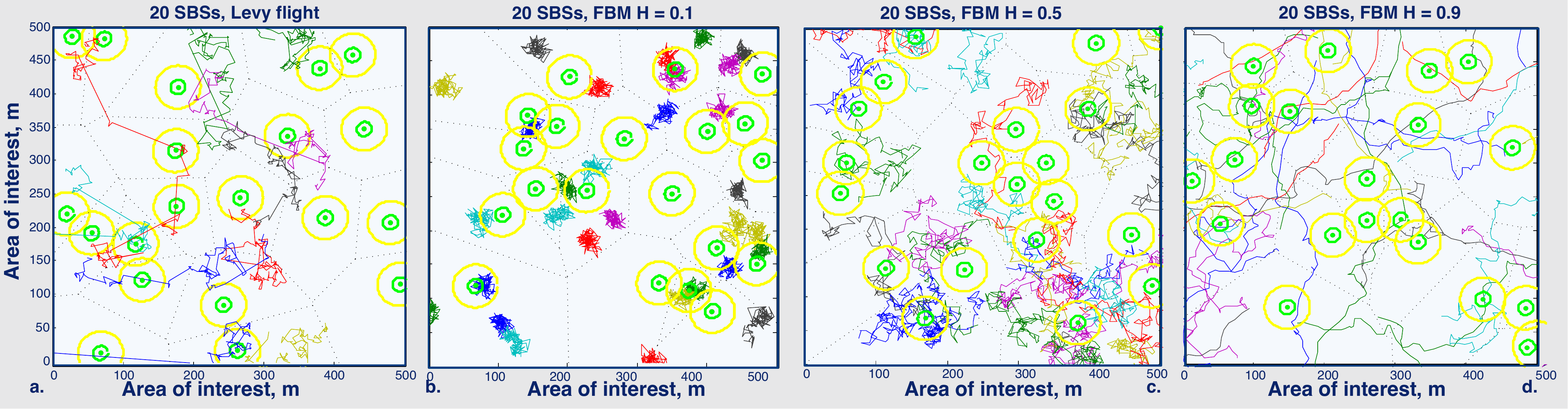}
\caption{User mobility trajectories (from left to right): (a) Levy flight, $\alpha=1.5$; (b) anti-persistent FBM, $H = 0.1$; (c) classical Brownian motion, $H = 0.5$; (d) persistent FBM, $H = 0.9$. }%Yellow and green circles indicate energy charging areas around SBSs for directional and omni-directional transmission, respectively. }
\label{fig:mobility}
\vspace{-0.6cm}
\end{figure*}

\section{Representative RF Power Transfer Scenario}

\subsection{Scenario Description}
%PLEASE, do not remove the comments below, they might be important in future

We consider WETI-enabled SBSs distributed following a Strauss process~\cite{Str75} over a particular area of interest. This process represents a model for spatial inhibition and ranges from Poisson point process to hard-core point process, according to the selected value of the interaction parameter. Further, \textcolor{black}{some of the main differences between energy and information transfer are} coverage discontinuity of the former dictated by lower receiver antenna gain, higher sensitivity threshold, and, consequently, much smaller coverage radius. \textcolor{black}{Therefore, it is crucial to capture the correlation between the consecutive locations of a user, which may be done by means of applying an appropriate mobility model.}

%$\gamma \in [0,1]$, respectively. While the completely random pattern is analytically tractable, hard core process is more realistic, since the deployed SBS s are separated by some distance.

In our scenario, users are initially distributed uniformly within the area of interest, and immediately start moving across this area according to the fractal Brownian motion (FBM)~\cite{Nor96} or Levy flight~\cite{Lee08} model (assuming "portals" for wrap-around along the borders). The self-similar FBM model with high Hurst parameter $H$ leads to a \textit{persistent} process, where the increments are normally-distributed as in conventional Brownian motion (BM), but no longer independent. While FBM includes correlations, as well as preserves the short tails of the normal distribution, Levy flight process has long tails, with uncorrelated increments. 

In Fig.~\ref{fig:mobility}, we provide an illustration of the SBS deployment together with the trajectories for the chosen user mobility models. Sub-figures (b) and (d) confirm high correlation between the sizes of increments, while sub-figures (a) and (c) demonstrate Gaussian coordinate changes and heavy-tailed step-length distribution, correspondingly. More specifically, Voronoi cells (dotted lines) illustrate the service areas of the SBSs; the coverage of directional energy transfer (yellow) is visibly wider than the omni-directional wireless transfer range (green) due to the beam directivity gains. 

%\begin{figure}
%\centering
%\includegraphics[width=0.7\columnwidth]{harvesting_walk2.pdf}
%\caption{Representative scenario}
%\label{fig:harvesting_walk}
%\end{figure}
%place somewhere: Refer to D. Niyato review! Many theoretical works in literature, but there is a gap between practice and...

Recall that the said energy coverage radius $R_E$ is calculated as the maximum distance, where received signal strength exceeds receiver sensitivity, which, in turn, depends on the antenna gain, the maximum power, and the path loss:
\begin{equation}
R_E = \sqrt{\frac{1}{S} p_{tx} G_{tx} G_{rx}} \left(\frac{\lambda}{4\pi}\right), \label{eqn:friis2}
\end{equation}
where $S$ is the rectifier sensitivity. %, $p_{tx}$ is the radiated power, $G_{tx}$/$ G_{rx}$ is the transmitter/receiver antenna gain, and $\lambda$ is the wavelength. 
Clearly, directional and omni-directional transmissions differ by their transmit power $p_{tx}$ and antenna gain $G_{tx}$, where the latter depends on the number of active antenna elements, that defines the width of the beam. For the directional antenna, the maximum allowed power is itself a decreasing function of antenna gain.% which should be considered by the regulatory documents.

%Independently of the selected directional/omni-directional scenario, when the user crosses the coverage radius, the associated devices begin receiving energy if the user body is not blocking the devices until the user leaves the coverage area. % (blocking is assumed to be happening with a certain probability, depending on the receiver antenna pattern)
In our setup, the minimum power of the received RF signal that enables wireless charging is set to $-20$ dBm, while the cumulative efficiency of the rectifier (RF-DC conversion) and the energy storage (DC to capacitor) equals $50\%$. In addition, the maximum power is decreased by $1$ dB for each $3$ dBi of antenna gain above $6$ dBi. Further, we assume that one user is associated with a single beam without (self-)blockage, and all user's wearable devices are located around the axis of the beam's symmetry. The collected power is thus calculated according to the Friis equation, and the energy radius is defined by the rectenna sensitivity, as shown by formula~(\ref{eqn:friis2}). 

For simplicity, all energy received from different SBSs is accumulated in the energy storage of a certain capacity, while the initial charge levels of wearables are uniformly-distributed. We also assume that a wearable device continues charging even if its energy storage is almost full due to constant discharge. The example values of carrier frequency, maximum radiated power, and energy coverage radius correspond to the column for $915$ MHz in Table I. Finally, user trajectories are characteristic of FBM (with varying parameter $H$) and Levy flight ($\alpha = 1.5$) patterns for $100$ users moving with the average speed of $3$ km/h. %The entire simulation is carried out in MATLAB environment.

\subsection{Understanding Performance Results}

To investigate the system-level performance of our envisioned RF charging service, we calculate (i) the share of effective time \textcolor{black}{(i.e., the average normalized device operating time, termed for brevity ANDOT in the figures below)} and (ii) the distribution of energy collected by a user per second considering both omni-directional and directional energy transfer.

%\begin{figure*}[!ht]
%\centering %
%\begin{tabular}{cc}
%\includegraphics[width=0.46\textwidth]{draft1}  & 
%\includegraphics[width=0.45\textwidth]{draft2}\tabularnewline
%\end{tabular}\protect\caption{Sample results [work in progress]}
%\label{fig:length} 
%\end{figure*}

\begin{figure*}[!ht]
\centering %
\vspace{-5px}
\begin{tabular}{cc}
\includegraphics[width=0.9\textwidth]{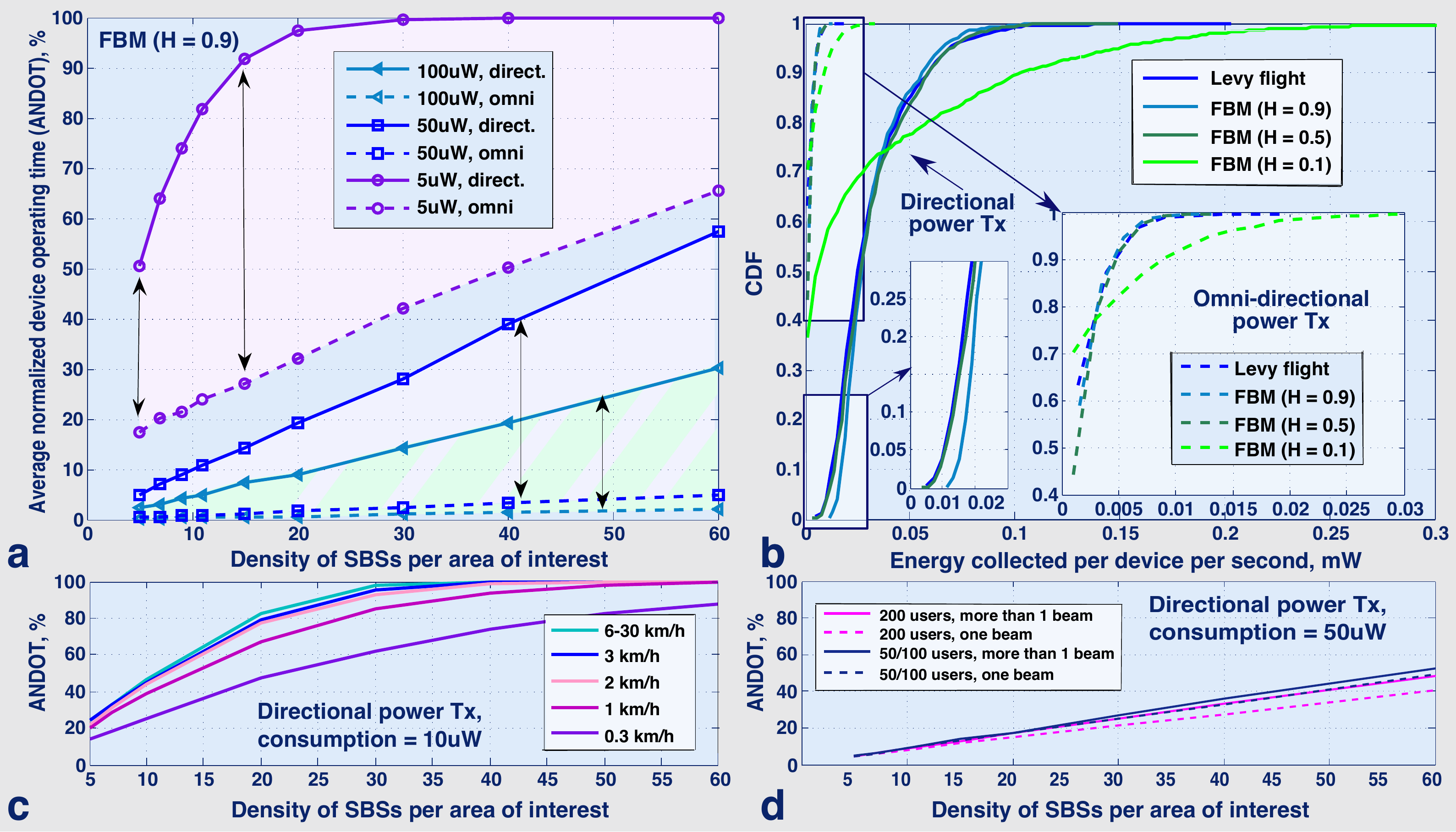} 
\end{tabular}
%\vspace{-5px}% \hspace{2px}
%\begin{tabular}{m{90px}}
%\includegraphics[width=0.4\textwidth]{pics/speed2}  & \hspace{5px}
%\includegraphics[width=0.40\textwidth]{pics/speed2}\tabularnewline
%\end{tabular}
\protect\caption{(a) Average share of active time (battery charge is above zero) for wearable devices with different energy consumption profiles under directional (solid) and omni-directional (dashed) charging regime; (b) CDF of energy collected by a device (on average, per second); (c) dependence on the user movement speed; (d) effects of user density.}
\label{fig:results2} 
\vspace{-0.6cm}
\end{figure*}

Collecting the values of the average operating time (i.e., time that wearables spend in active mode) for three different energy discharge rates, Fig. \ref{fig:results2}(a) illustrates the evolution of the directivity gain with the increasing SBS density. Depending on the energy consumption profile, the said gain may be very promising (see $5\mu W$ with up to $60\%$ of absolute increase for $15$ SBSs). After a certain SBS density (e.g., $30$ SBSs per area of interest $500$x$500m^2$ as indicated by $5\mu W$ solid line), wearable devices with lower energy consumption can operate supported solely by RF charging. As long as the SBS density remains low, the observed trends are near-linear, whereas for higher densities the dependencies assume exponential form.

Further, Fig. \ref{fig:results2}(b) illustrates the distribution of energy collected by a user for directional (dashed lines) and omni-directional (solid lines) RF power transmission. In particular, the "omni-directional" results, as well as the lower-left corner of the "directional" plots, are enlarged for convenience. Here, the two extremes are the realizations of the FBM with the parameters $H = 0.9$ and $H = 0.1$, where the former provides a more fair energy charge distribution, while the latter leads to a more diverse picture, since the users are not moving much. The performance of the Levy flight model is fairly close to the BM case (e.g., FBM with $H = 0.5$) due to the absence of correlation between two consecutive increments. 

For any considered mobility model, the comparison of cumulative distribution functions (CDFs) for directional and omni-directional energy transmission indicates clear superiority of the former regime, which generally confirms our expectations. \textcolor{black}{Importantly, the results reported in Fig. \ref{fig:results2}(b) do not depend on the energy consumption profile, since they are measured only when the energy is being received.} We note, however, that our provided estimates are tentative, and the ultimate system performance will largely depend on the official RF energy transfer regulations for allowed radiated power.

Finally, for directional power transmission with fixed energy consumption, the shape of the aforementioned parameter (ANDOT) with respect to the average speed and the density of users, respectively, is illustrated in Fig. \ref{fig:results2}(c) and (d). Accordingly, after a certain user density, the system performance decreases slightly for any SBS density. The same effect may be observed, when the user is served by only one, its closest SBS (see "one beam" in the legend). Similarly, for some user speeds, the average active time per device may drop due to less user activity and longer cell residence times.

In summary, our performance evaluation suggests attractive benefits after incorporating the dedicated RF charging capability into the 5G ecosystem. However, there remain certain challenges in unlocking these promising benefits, that embrace a range of aspects, from technical and economical issues to more advanced system modeling. 

%%%-------------------------------------------------------------------------------------------------------------------------------------------------------------------------------------------------------------------------------------------------------------------------------------------------------------------------------------------------------------------- 
 
\section{Future Challenges and Open Issues}

%Having started with a preliminary system-level evaluation, which concentrated on the interaction between the SBSs and the wearables they charge, the effects of user mobility, and the spatial component, 
In this section we continue by focusing on the system modeling challenges and elaborate on the new problem formulations, specific for the RF power transfer.

\subsection{Technology-Related Challenges}\label{DirectivityModellingChallenges}

\begin{figure}[!ht]
\centering
\includegraphics[width=0.95\columnwidth]{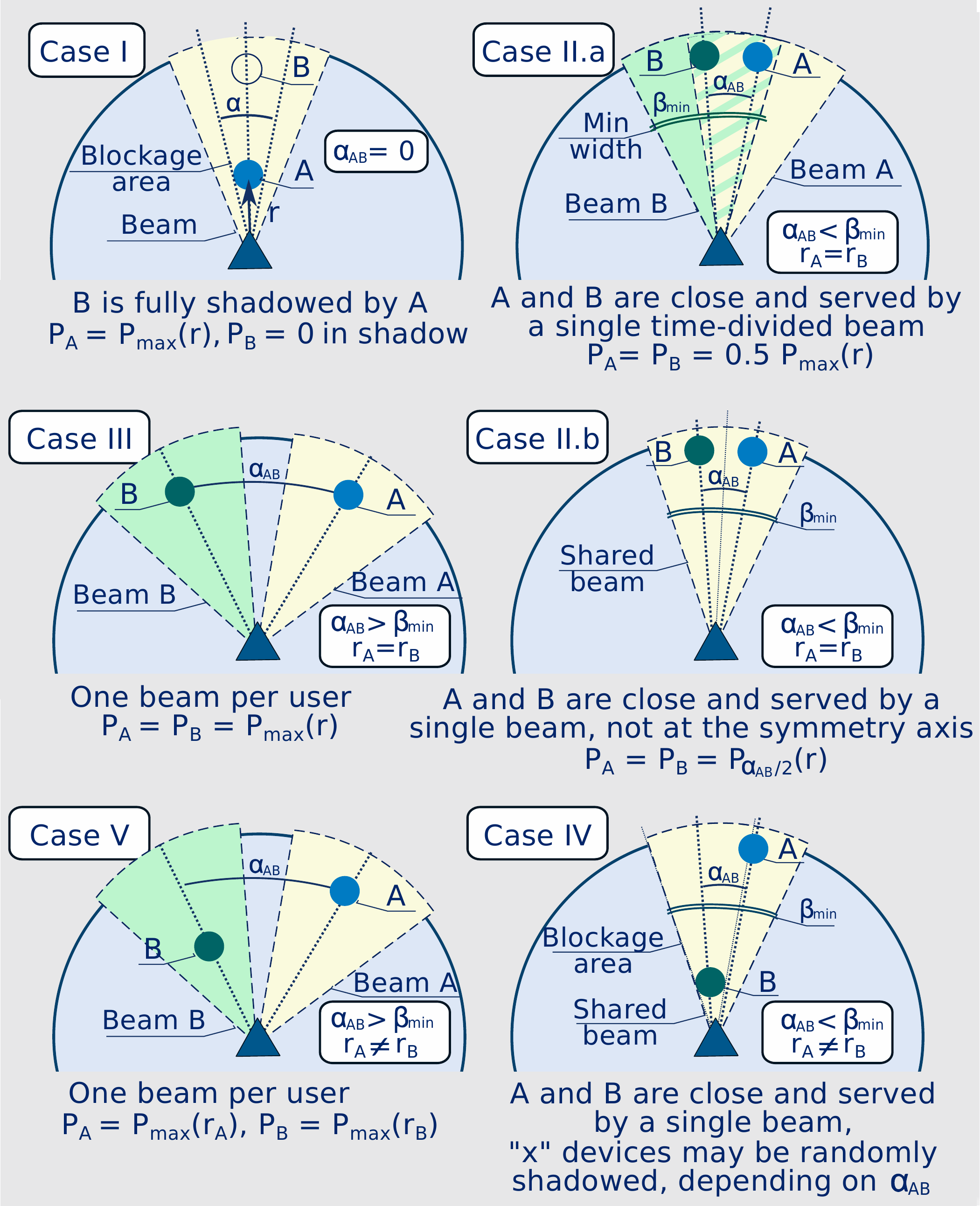}
\caption{Directional charging scenarios. $r_{A/B}$ is the radial coordinate of user A/B, $\alpha_{AB}$ is the angular distance between the users, $\beta_{\min}$ is the minimum beam width, which defines the angular resolution allowing to direct the beam so that the user will be located on its axis of symmetry.}
\label{fig:layout}
\vspace{-0.6cm}
\end{figure}

%While omni-directional RF energy transfer , directional transmission has its own unique challenges. 
In reality, wearable devices are clustered on their users (potentially including adults, kids, and pets) and within the SBS energy coverage may be charged simultaneously if there is no self-blockage. To classify the possible options of relative user locations, we consider two users in the proximity of one power source, which may easily be extrapolated for the case of multiple users and sources. Referring to Fig. \ref{fig:layout}, we provide below a description of several characteristic scenarios with different angular and radial coordinates of users.

%(More detailed discussion on the directional transmission)
\subsubsection{Case I (angular distance is zero)}
One user is fully shadowed by another. The devices of the closest user collect power according to the Friis law, while the shadowed wearables acquire close to no energy from the power source. 
%Or one client is placed exactly between the second client and the base station thus blocking the line of sight. In such a case we can either assume that the second client is not getting any power even though he is
%covered by the beam.

\subsubsection{Case II (equal distances to the SBS, angular distance is less than the beam width)}
The following options are possible, and two users may be served by: 
(a) a single time-divided beam, when directional charging alternates from one user to another (received energy per second is therefore reduced proportionally, but either user may receive some additional energy from the beam directed at the neighbor, depending on the beam width),
(b) a single beam, when the users are not at the symmetry axis (the received energy is decreased due to the respective angle). 

If the user shifts from the axis of symmetry by the angle $\varphi$, the received power is decreased by approximately $|\sin(N\varphi/2)/N\sin(\varphi/2)|^{2}$, where $N$ is the number of antenna elements (adopted from~\cite{Orf14}, (20.7.4)). Two neighboring users may also be served by increasing the beam width, while at the same time reducing the antenna gain. The flexibility of the choice here spawns a set of practical questions on how the actual scheduling and beamforming should be performed to guarantee the most effective service.

\subsubsection{Cases III, V (angular distance is large)}
Each user is served by an individual beam as long as the number of simultaneous beams does not exceed the maximum following e.g., from the total radiated power constraints. The available power is then defined by the corresponding distance.

\subsubsection{Case IV (different distances to the SBS, short angular distance)}
Two users are relatively close to each other and may be served by a single (time-divided) beam. However, some wearable devices might be randomly shadowed, depending on the angular distance between the users and the current position of their devices. Hence, accurate characterization of the beam-blockage probability becomes a challenging problem, especially in light of extreme diversity in sizes, locations, and shapes of obstacles. 
%\subsubsection{Case V (different distances to the BS, )}
%Two users are relatively far from each other, different distance from BS. One beam per user.

 %\subsection{System-level challenges}
Generalizing the above to higher densities of SBSs and users, the effective constraints on the maximum number of beams per antenna and the respective transmit power limit per beam become even more critical, and thus have to be modeled comprehensively. Moreover, future RF charging networks might comprise of a variety of energy sources equipped with very diverse capabilities for energy transfer. Another open problem is that related to charging frequency selection, where the uncontrolled use of ISM bands may cause severe unpredictable interference, which has to be taken into account. Finally, more complex beam scheduling procedures for practical overcharging policies may be required, which service operators can introduce in their wireless charging systems to balance energy replenishment for wearables with different charge levels.

%%%--------------------------------------------------------------------------------------------------------------------
%%%--------------------------------------------------------------------------------------------------------------------
%%%--------------------------------------------------------------------------------------------------------------------
%%%--------------------------------------------------------------------------------------------------------------------
%%%--------------------------------------------------------------------------------------------------------------------
\vspace{-0.2cm}
\subsection{Analytical Modeling Challenges and Solutions}

The technical challenges and differences between the data and energy transmission give rise to multiple open problems also from the system analysis point of view, which have to be well articulated. In this subsection, we provide our vision on the limitations of existing models as commanded by the technical challenges.

\subsubsection{Beamforming-related issues}
In particular, when the number of users in the area of interest is large due to less sensitivity (i.e., wider coverage radius) or high user density, the following issues arise:
\begin{itemize}
\item {\em Reception of Multiple Overlapping Beams:} where the chances of a given user to receive energy from multiple directional beams (i.e., the beams that are intended to serve the neighbors of the considered user) become high.  In such a case, the existing theoretical frameworks consider  the energy acquisition only from the intended main beam of a user. This provides a worst-case estimate on the harvested energy and has to be refined by the respective modeling. 
\item {\em Blocked Desired Beams:}
On the other hand, considering the orientation of the directional energy beam, two or more users may co-exist within the same beam and, in turn, their received power may become partially or fully blocked. This requires a careful modeling of the beam blockage events as a function of the density of the SBSs and users, mobility patterns, and orientation of users. Further, to reduce the probability of beam-blockage and thus energy outage events, new user scheduling/offloading mechanisms and SBS coalition-formation strategies need to be developed. 
\end{itemize}

Moreover, with a higher number of devices in the vicinity of an energy provider, the probability of energy outage due to the limited number of beams and transmit power per beam would increase. To overcome this and the above issues, the existing analytical models need to be modified to reflect accurately the relation between the number of beams, their allowed maximum transmit power, and the number of charging devices that can be served reliably. %Furthermore, novel energy-aware user scheduling and clustering schemes would play a crucial role in the successful energy transfer implementation.

For the purposes of analytical tractability, the actual antenna patterns may also be approximated by a sectored antenna that includes the key features of an antenna pattern, such as the main lobe directivity gain, which is a function of the angle of arrival and the angle of departure, side/back lobe gains, beam width of the main lobe, and the angle of the boresight direction. In the sectored antenna model, the array gains are taken as constants for all angles in the main lobes and the side lobes. Stochastic ordering tools can then be utilized to comparatively analyze and optimize the amount of harvested energy as a function of different parameters of the antenna pattern.

%\subsubsection{Distribution of the Effective Beams per User}
Finally, given that a user may sense the directed unblocked energy intended for other users in a certain surrounding region, it is crucial to accurately model the distribution of the number of effective beams and their strength within a particular area around a given user and, importantly, their availability for the associated wearable devices. While for the static/snapshot models assessing the distribution of the number of users can be straightforward, determining it as a function of mobility patterns may become a challenging task.

\subsubsection{Mobility Modeling in Wireless Charging Networks}

As argued above, mobility may become a crucial factor in system-level analysis of RF power transfer. In this context, user mobility may sometimes be captured analytically via cell or zone residence time (defined as the time spent by a user within the SBS coverage or the charging zone, respectively). The overlap between technologically-different energy transfer zones leads to a correlation between the zone residence time (ZRT) and the cell residence time (CRT). The time spent in each cell/zone may be modeled as a generic \textit{Phase-Type (PH) distribution} due to its specific properties of~(i) approximating any non-negative statistics and (ii)~a closure property, by which several operations on PH distributions produce another PH distribution.

The correlation between the CRT and ZRT can be represented by defining CRT as a stochastic sum of ZRTs. Traditionally, for queuing systems, the Coxian structure is proposed, where customers probabilistically visit different numbers of servers and then exit. The same concept can in principle be applied here, since durations spent in different zones are analogous to the time spent at different servers, and the cell exit probabilities are analogous to the system exit probabilities. From the user mobility characterization, other important variables follow immediately, equally deserving research attention, such as e.g., dynamics of charging sessions. This process is affected by the energy consumption profile and can be characterized, for example, by different absorption states. That is, normal harvesting session termination (upon the desired level of battery charge), successful hand-off to a neighboring cell/zone, session drop during a cell/zone hand-off, etc.

%%%--------------------------------------------------------------------------------------------------------------------
%%

\section{Conclusions and Lessons Learned}

Our proposed vision of dedicated RF charging capability opens door to "self-sustainable" wearable solutions. Hence, it brings along transformative changes to future 5G deployments by enabling autonomously-powered wearable devices without high cost for the network. In addition to enjoying the reduced use of conventional energy, next-generation wireless-powered wearables promise their users the untethered, true mobility thus making the feasible RF powering technology a new commodity. To this end, our present study reveals the following main findings:

1) Integrated RF power chargers on top of the multi-radio SBSs might become a promising 5G-grade solution.% due to reduced CAPEX/OPEX.

2) Coexistence of operator-deployed and user-deployed SBSs leads to profoundly different incentivization and energy sharing mechanisms.

3) Out-of-band SWIPT (possibly, in bands of around $800-900$ MHz) allows for more flexible receiver architectures and, therefore, may be preferred at the initial stages of system implementation. However, unpredictable interference in ISM bands has to be considered.

4) Owing to contemporary cellular-assistance functionality, directional energy transmission is promising due to its higher efficiency, which has been confirmed by our system-level evaluation. However, the power consumption of wearables has to be reasonably low \textcolor{black}{(less than a milliwatt)} for the wireless RF charging to remain effective.

5) Due to expected discontinuous coverage of dedicated RF power sources, users traveling between the available wireless chargers more actively might benefit further. In turn, user density may have a pronounced effect on the choice of the appropriate scheduling and beamforming solutions.

\section*{Acknowledgment}
\thanks{This work is supported by the Academy of Finland. The work of the fourth author is supported with a Postdoctoral Researcher grant by the Academy of Finland as well as with a Jorma Ollila grant by Nokia Foundation.}
 
\vspace{-0.3cm}
\bibliographystyle{ieeetr}
\bibliography{refs}

\clearpage
\section*{Authors' Biographies}

\textbf{Olga Galinina} (olga.galinina.tut.fi) received her Ph.D. degree from the Department of 
Electronics and Communications Engineering at Tampere University of Technology, Finland. She previously received her B.Sc. and M.Sc. degrees in Applied Mathematics from the Department of Applied Mathematics, Faculty of Mechanics and Physics, St.-Petersburg State Polytechnical University, Russia. %She has publications on mathematical problems in the novel telecommunication protocols in internationally recognized journals and high-level peer-reviewed conferences. 
Her research interests include queueing theory, stochastic processes, optimization theory and their applications; heterogeneous wireless networking, machine-to-machine, wearable, and device-to-device communications.

%\textbf{Hina Tabassum} (M'13)
%received the B.E. degree in electronic engineering from the NED University of
%Engineering and Technology (NEDUET), Karachi, Pakistan, in 2004. She received during her undergraduate studies 2 gold medals from NEDUET
%and SIEMENS for securing the first position among all engineering universities of Karachi. She then worked as lecturer in NEDUET for two
%years. In September 2005, she joined the Pakistan Space and Upper Atmosphere Research Commission (SUPARCO), Karachi, Pakistan and
%received there the best performance award in 2009. She completed her
%Masters and Ph.D. degree in Communications Engineering from NEDUET
%in 2009 and King Abdullah University of Science and Technology (KAUST),
%Makkah Province, Saudi Arabia, in May 2013, respectively. Currently, she
%is working as a post-doctoral fellow in the University of Manitoba (UoM),
%Canada. Her research interests include wireless communications with focus on interference modeling, spectrum allocation, and power control in
%heterogeneous networks.

\textbf{Hina Tabassum} (hina.tabassum@umanitoba.ca) received her B.E. degree in electronic engineering from the NED University of Engineering and Technology (NEDUET), Karachi, Pakistan, in 2004. During her undergraduate studies she received two gold medals from NEDUET and Siemens for securing the first position among the students of all the engineering universities in Karachi. She then worked as a lecturer at NEDUET for two years. In September 2005, she joined the Pakistan Space and Upper Atmosphere Research Commission (SUPARCO), Karachi, Pakistan and received there the best
performance award in 2009. She completed her Master's and Ph.D. degrees in communications engineering from NEDUET in 2009 and King Abdullah University of Science and Technology (KAUST), Makkah Province, SaudiArabia, in May 2013, respectively. Currently, she is working as a post-doctoral fellow at University of Manitoba (UoM). Her research interests include wireless communications with focus on interference modeling and performance analysis, spectrum allocation, energy harvesting, and full-duplex communications in heterogeneous cellular networks.

\textbf{Konstantin Mikhaylov} (konstantin.mikhaylov@ee.oulu.fi) received his B.Sc. (2006) and M.Sc. (2008) degrees in Electrical Engineering with the focus on wireless systems from St. Petersburg State Polytechnical University, St. Petersburg, Russia. Since 2009, he works as a researcher for wireless and embedded systems in the University of Oulu, Oulu, Finland. His research interests include resource constrained embedded systems, energy-efficient wireless communication technologies, wireless sensor and actuator networks and their practical applications. Over the recent years, he has participated in multiple industrial projects, authored and co-authored over 30 research and technical papers focusing on the different aspects of energy-efficient wireless communications.
       
\textbf{Sergey Andreev} (sergey.andreev@tut.fi) is a Senior Research
Scientist in the Department of Electronics and Communications
Engineering at Tampere University of Technology, Finland. He received
the Specialist degree (2006) and the Cand.Sc. degree (2009) both from
St. Petersburg State University of Aerospace Instrumentation, St.
Petersburg, Russia, as well as the Ph.D. degree (2012) from Tampere
University of Technology. Sergey (co-)authored more than 90 published
research works on wireless communications, energy efficiency,
heterogeneous networking, cooperative communications, and
machine-to-machine applications.

\textbf{Ekram Hossain} (ekram.hossain@umanitoba.ca) is a Professor in the Department of Electrical and Computer
Engineering at University of Manitoba, Winnipeg, Canada. He received his Ph.D. in Electrical
Engineering from University of Victoria, Canada, in 2001. Dr. Hossain's current research
interests include design, analysis, and optimization of wireless/mobile communications networks, cognitive and green radio systems, and network economics. He has authored/edited several books in these areas (http://home.cc.umanitoba.ca/$\sim$hossaina). He was elevated to an IEEE Fellow ``for contributions to spectrum management and resource allocation in cognitive and cellular radio networks".  Currently, he serves as the Editor-in-Chief for the \textit{IEEE Communications Surveys and Tutorials} and an Editor for \textit{IEEE Wireless Communications}. Also, he is a member of the IEEE Press Editorial Board. Previously, he served as the Area Editor for the \textit{IEEE Transactions on Wireless Communications} in the area of "Resource Management and Multiple Access" from 2009-2011, an Editor for the \textit{IEEE Transactions on Mobile Computing} from 2007-2012, and an Editor for the \textit{IEEE Journal on Selected Areas in Communications - Cognitive Radio Series} from 2011-2014. Dr. Hossain has won several research awards including
the IEEE Communications Society Transmission, Access, and Optical Systems (TAOS) Technical Committee's Best Paper Award in IEEE Globecom 2015, University of Manitoba Merit Award in 2010 and 2014 (for Research and Scholarly Activities), the 2011 IEEE Communications Society Fred Ellersick
Prize Paper Award, and the IEEE Wireless Communications and Networking
Conference 2012 (WCNC'12) Best Paper Award. He is a Distinguished Lecturer of the
IEEE Communications Society (2012-2015). He is a registered Professional
Engineer in the province of Manitoba, Canada.

\textbf{Yevgeni Koucheryavy} (yk@cs.tut.fi) is a Full Professor and
Lab Director at the Department of Electronics and Communications
Engineering of Tampere University of Technology (TUT), Finland. He
received his Ph.D. degree (2004) from TUT. He is the author of
numerous publications in the field of advanced wired and wireless
networking and communications. His current research interests include
various aspects in heterogeneous wireless communication networks and
systems, the Internet of Things and its standardization, as well as
nanocommunications. He is Associate Technical Editor of IEEE
Communications Magazine and Editor of IEEE Communications Surveys and
Tutorials.

\end{document}